# Searching by index for similar sequences: the SEQR algorithm

David I. Hurwitz, Lianyi Han, and Lewis Y. Geer[1]

National Center for Biotechnology Information, National Library of Medicine, National Institutes of Health, Bethesda, Maryland, USA.

## Abstract

This paper describes a method to efficiently retrieve protein database sequences similar to a query sequence, while allowing for significant numbers of mutations.  We call this method SEQR for **SEQ**uence **R**etrieval.  This approach increases the speed of sequence similarity searches by an order of magnitude compared to conventional algorithms at the expense of sensitivity. Furthermore, retrieval time increases less than linearly with the number of sequences, a desirable property during an era when next generation sequencing technologies have yielded greater than exponential increases in sequence records. The lower sensitivity of the algorithm for distantly related sequences compared to benchmarks is not intrinsic to the method itself, but rather due to the procedure used to construct the indexing terms, and may be improved. The indexing terms themselves can be added to standard information retrieval engines, enabling complex queries that include sequence similarity and other descriptors such as taxonomy and text descriptions.

## Introduction

The number of protein sequences has grown from the 65 contained in the Atlas of Protein Sequence and Structure in 1965 (Dayhoff, Eck et al. 1965) to over 60 million today, (EMBL (2016), NCBI (2016)) growing at a more than exponential rate for long periods (Wetterstrand (2016)).  Finding similar sequences, an essential step in determining the function of a query sequence, has progressed from inferring distant homologies from a database of only a few sequences to sorting through large sets containing many similar sequences. Popular sequence search algorithms, such as HMMER (Eddy 1995) and BLAST (Altschul, Madden et al. 1997), linearly search through sequence databases, with a computational cost that can exceed the exponential improvement in computational hardware over time. In this study we explore an algorithm that searches for similar sequences by using inverted indices. Search time for this method increases less than linearly with database size, and can approach $O(\log(n))$ growth, depending on the retrieval method selected.  Additionally, since inverted indices are a standard retrieval method, the similarity indices may be combined with other indices, enhancing the ability to find similar sequences with desired functional annotation.

A key step in inverted index retrieval is the generation of indexing terms.  A common technique is to compute indexing terms based on descriptors calculated from the records themselves. Examples of this technique include chemical similarity searching (Willett, Barnard et al. 1998) and image retrieval (Kato 1992). A chemical structure might be indexed by the functional groups or the common subgraphs it contains, while an image might be indexed by color or morphological elements of the image. In the case of sequences, indexing by exact or nearly exact sequence n-mers has shown high search efficiency but at the loss of sensitivity to mutations (Kent 2002) (Ning, Cox et al. 2001) (Li and Godzik 2006). Variants of exact match searches use amino acid reduced alphabets to allow for mutations (Edgar 2010) while clustering methods using similar k-mers show significant performance benefits (Hauser, Mayer et al. 2013).

---

[1] Email: lewis.geer@gmail.com

In the method described in this study, mutations are allowed by creating indexing terms based on discrete Position Specific Scoring Matrices (PSSMs) (Stormo, Schneider et al. 1982), which are statistical profiles of the amino acids found at various positions in sequences. These statistical profiles describe mutation rates at particular sequence positions and are created using mutations rates found in nature. We derive our indexing PSSMs from standard substitution matrices used to score sequence alignments for significance. Each of these indexing PSSMs is assigned a unique identifier. Indexing a sequence is simply a matter of collecting the unique PSSM identifiers for each n-mer in the sequence. Creating inverted indices for retrieval is straightforward: each sequence in the search database is indexed, and if a sequence contains an indexing term, the sequence identifier is stored with that indexing term. Upon query, the indexing terms for the query sequence are used to retrieve the corresponding inverted index, and the inverted indices found are used to retrieve the sequence records containing the indexing terms in the query.

Generation of the indexing PSSMs can take many forms and the following study examines several permutations to find optimal retrieval efficiency. We compare the resulting search to DELTA-BLAST (Boratyn, Schäffer et al. 2012), a commonly used algorithm. In the course of this comparison, various scoring methods are examined. Finally, the flexibility of this retrieval method may lead to various configurations of searches, and we examine several in pursuit of higher sensitivity.

## Methods

### Overview

Our method creates clusters of similar n-mers, then uses these clusters as indexing terms for database indexing and fast sequence retrieval. Each sequence in a protein database is indexed by converting each n-mer in the sequence to its corresponding cluster index. The indexed sequences are saved in a database. Query sequences are indexed as well, and their index terms are used to query the database for sequences with matching terms. A popular search platform, Apache Solr, is used for fast sequence retrieval, although the algorithm may be implemented using any standard information retrieval engine. Figure 1 shows an overview of the method.

For the sequence retrieval to be as sensitive and specific as we can make it, we optimized several parameters when making the n-mer clusters. For example, we varied the following: the substitution matrix used when making the n-mer clusters, the threshold for grouping n-mers into clusters, and the number of residues in the n-mer. Other parameters were adjusted and are discussed below.

To optimize these parameters, we ran many experiments in which we queried an indexed database with index terms from a random set of about 300 query sequences. In each case, we compared our results against a set of "true-positives" we had made by running DELTA-BLAST on the same query sequences. We judged the performance of our method by making ROC curves for each combination of parameters, and chose the parameters that gave us the most sensitive and specific curves, paying particular attention to the high-threshold portion of the curve where users are generally most interested.

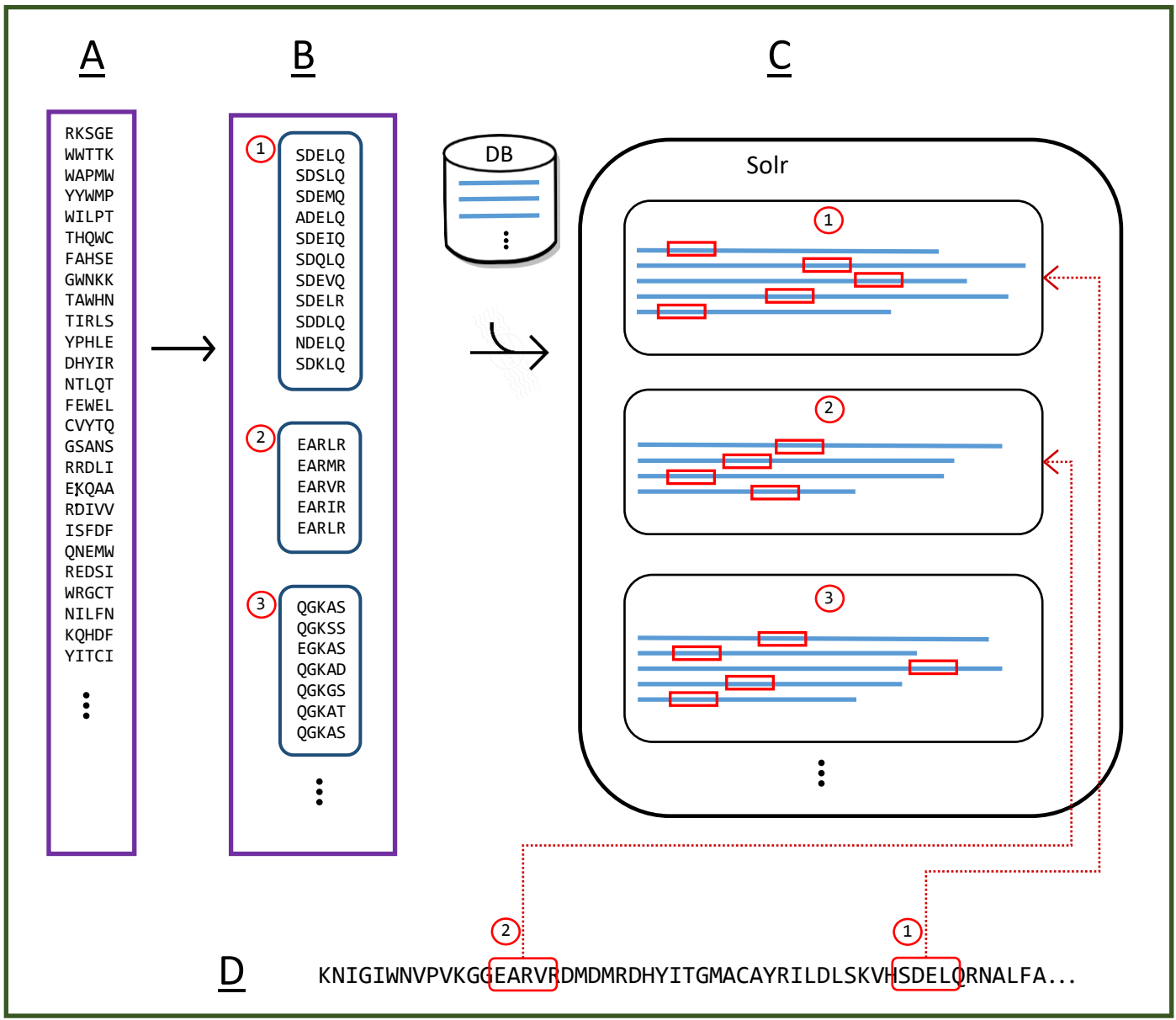

**Figure 1.** This figure shows the overall method for clustering nmers and indexing a database of sequences. From left to right: 3.2 million randomly shuffled 5-mers **(A)**, are clustered into groups of closely related n-mers **(B)**. Each cluster is assigned an integer identifier. These identifiers are the indexing terms. Next, a database of sequences is indexed using these clusters. Each sequence is traversed, and the cluster index at each position is determined. The cluster indices for each sequence are stored in a database using inverted indices so that each indexing term refers back to the sequences that contain each term **(C)**. We use Apache Solr to store the database of sequences. When a query sequence accesses the database **(D)**, the query sequence is also indexed, and the database can quickly find the sequences that contain each indexing term. The sequences are returned from the database sorted so that the sequences with the most matching terms are returned first. We have configured Solr to sort the hits based on a Jaccard index.

## Making indexing PSSMs

The basic unit of the SEQR algorithm is a cluster of similar n-mers, each assigned an indexing ID. Each cluster consists of a unique integer ID, a set of similar n-mers (one of which is a seed), a seed PSSM, and an average PSSM.

Creating these clusters was initially done via single linkage clustering. Later we used a fast-clustering method outlined below. In both methods, we avoid selecting n-mers in lexicographic order as this runs the risk of creating unbalanced clusters. Instead, the full set of n-mers was enumerated and randomly shuffled at the outset, then the shuffled n-mers were used. We used a different random number seed each time we clustered n-mers.

For single linkage clustering, the shuffled n-mers are compared to a growing list of n-mer clusters. If the comparison score is above threshold, the n-mer is added to the cluster. Otherwise, the n-mer is used to initiate a new cluster. In this work we tested different cluster thresholds, different BLOSUM (Henikoff and Henikoff 1992) and PAM (Dayhoff, Schwartz et al. 1978) substitution matrices for making the n-mer to cluster comparisons, different n-mer lengths, and whether to add n-mers to the first or best matching cluster in the list. The results of these tests and others are described below.

After clustering completes, each cluster is assigned an integer identifier. The process can stopped at this point, or the clustering can be iteratively refined by reassigning all n-mers to the existing clusters.

When this process is complete, the cluster's integer ID becomes the indexing term that is applied to all n-mers in the cluster. This simple relationship allows the creation of an index where the key is a sequence n-mer and the value is the indexing term. A simple method for creating this index is to lexicographically sort the n-mers list along with their corresponding index terms. This index can be used to look up the indexing term for a particular sequence n-mer.

## N-mer to cluster similarity

The distance between an n-mer and an existing cluster can be computed many different ways, of which we examined two. We either compared an n-mer to the PSSM of the seed of the cluster, or compared an n-mer to an average PSSM representing all n-mers in the cluster.

The PSSM of the seed of a cluster is created by iterating through each position of the seed's n-mer and placing the column for the corresponding amino acid from a BLOSUM or PAM mutation matrix into the appropriate PSSM column of the cluster. In other words, the PSSM of the seed of the cluster is composed of columns from a standard substitution matrix. The average PSSM of the cluster is created by making PSSMs for each n-mer in the cluster in the same manner, and averaging the PSSMs together. An example of a cluster's seed PSSM and average PSSM are shown in this paper's supplemental materials. Each time a new n-mer is added to a cluster, the PSSM for that n-mer is averaged into the cluster's average PSSM. Since the average PSSM of a cluster changes as new n-mers are averaged into this PSSM, the centroid of the cluster drifts during clustering. In contrast, there is no drift of the PSSM of the seed of the cluster since it is simply the PSSM of the first n-mer in the cluster.

N-mer to cluster scoring is performed in a manner similar to comparing 2 n-mers using a substitution matrix. The score for each residue to cluster-position is read from the cluster's substitution matrix, and then summed over the n-mer. Since both the PSSM of the seed and the average PSSM of the cluster are stored with the cluster, one can compare an n-mer to either the seed or the average PSSM of the cluster.

## Cluster-to-cluster similarity

In contrast to the n-mer to cluster comparisons made during clustering, during the 2-query method discussed below, clusters are also compared to other clusters. When calculating a similarity score for 2 clusters, we make the comparison by treating the seed of each cluster as an n-mer and comparing each seed to the PSSM of the other cluster. These 2 similarity scores are then averaged. As with the n-mer to cluster comparisons, each seed can be compared to either the PSSM of the seed of a cluster or the average PSSM of the cluster.

For the 2-query method, all cluster neighbors are pre-computed. The cluster-cluster comparisons that are above a cluster-neighbor threshold are saved in the cluster index file along with their similarity scores. Therefore, in addition to containing the cluster assignment for each n-mer, this file also contains the clusters that are closely related to each other cluster.

## Fast clustering algorithm

For this work, we developed a fast clustering method that makes it much more tractable to work with 5-mers and 6-mers. We did not cluster 7-mers or larger as clustering all possible 1.28 billion 7-mers would require significant algorithmic improvements or code parallelization.

The motivation for developing a fast clustering method is that our single-linkage clustering method, outlined above, requires several days of computational time for 5-mers with high thresholds, and weeks or months of computational work for 6-mers. These tests were performed on a single 2.2 GHz Intel Xeon machine. A fast clustering method allows 3.2 million 5-mers to be clustered in a few hours, and 64 million 6-mers to be clustered in days instead of weeks or months.

The underlying idea is that after running single-linkage clustering for a period of time, the known clusters are fully enumerated in a fast step. That is, all n-mers that are within threshold of each cluster's PSSM are quickly computed. After this step an n-mer will typically appear in the enumerated lists of multiple clusters. When we continue single linkage clustering, we can quickly see if a new n-mer belongs to any of the known clusters' enumerated sets. If yes, we compare the n-mer to the clusters in which the n-mer is found, and assign it to the highest scoring match. If no, we test the new n-mer against the clusters that were added since the enumeration step, and initiate a new cluster if no match is found. Either way, we save most of the computational work of comparing a new n-mer against all PSSMs.

The key to making the cluster-enumeration step fast is to observe that once we know that an n-mer to cluster comparison has a score below threshold, we also know that any n-mer that scores less than this will also score lower than threshold. Arranging each column of a PSSM from hi to low score allows us to cycle through the n-mer permutations in each column in decreasing score order. When we discover an n-mer whose comparison with the PSSM is lower than threshold, we can skip the remainder of the column. This allows us to skip most n-mer to PSSM comparisons during the enumeration step. In practice, it proved sufficiently fast to cycle through all n-mers and simply advance the fastest cycling column to the end when an n-mer scored below threshold. For longer n-mers, it may be necessary to make additional speed improvements to this step.

The work proceeds in rounds, alternating between adding n-mers during single-linkage clustering, then fully enumerating the new clusters. When the PSSM is allowed to drift, the enumeration step needs to be repeated after each round of adding n-mers to clusters, for any clusters that had new n-mers added to them, since the addition of new n-mers to a cluster will affect its average PSSM and, hence, the n-mers that are within threshold of the PSSM. Fast clustering is more similar to the best-hit method mentioned above, than the first-hit method, as each new n-mer is compared to all clusters that contain the n-mer in their enumerated sets.

## 2-query SEQR

We experimented with different methods to increase the sensitivity of SEQR, including using the results from an initial SEQR query to form a 2nd query. The aim is for the $2^{nd}$ query to pull in a wider range of hits than the $1^{st}$ query alone. The best of our 2-query methods is described here.

First, a SEQR query is done in the usual manner: the index terms of the query sequence are used to retrieve sequences from Solr. Sequences with a Jaccard index similarity to the query above a threshold are then used to form a $2^{nd}$ query. The threshold we used at this point was pre-determined, and was the point in a comparable 1-query ROC curve at which the number of true-positives and false-positives were roughly the same. This insured that most of the sequences used to make the $2^{nd}$ query were true-positives.

Next, the algorithm looks through these hits above threshold from the $1^{st}$ query to find the most popular index terms. When looking for frequently-occurring index terms, we do not consider all possible index terms. Rather, we consider only the index terms from the query and their pre-computed cluster-neighbors. A cluster-neighbor is a cluster of n-mers that have a similarity-score with another cluster above a threshold. (See the section above on "Cluster-to-cluster similarity"). These frequently-occurring index terms, that are either index terms of the query, or cluster-neighbors of the query's index terms, are used to make the $2^{nd}$ query.

The purpose of the $2^{nd}$ query is to allow for variations of the index terms from the $1^{st}$ query. These variations are limited to those frequently found in the hits of the $1^{st}$ query. Allowing all frequently-occurring index terms from the hits of the $1^{st}$ query to be used in the $2^{nd}$ query results in poorer retrieval [results not shown], perhaps because index terms from adjacent protein domains not found in the query are pulled into the $2^{nd}$ query this way.

Finally, we pool all the hits from the 2 queries and score them. We found that if we re-score all the hits from the 2 queries on an equal footing we get the best results in terms of sensitivity and specificity. To do this, we used a modified Jaccard index similarity score to compare all the pooled hits to the $2^{nd}$ query. In contrast to the Jaccard index score discussed below, the score here not only counts exact index matches between the query and a hit, but also counts matches of an index term to one of its cluster neighbors. In this modified Jaccard index, the number of matching index terms is the number of index terms in the hit that match either a query's index terms or the index terms of one of its cluster-neighbors used in the $2^{nd}$ query.

During our 2-query testing, we experimented with varying the cluster-threshold, the cluster-neighbor-threshold, and the number of index terms as a fraction of the number of index terms in the original query used in the $2^{nd}$ query. The results are presented below.

## 1.5-query SEQR

During the course of developing our 2-query SEQR method, we discovered that if we run all the steps of 2-query SEQR *except* querying Solr a $2^{nd}$ time, our ROC curves are improved over running normal 1-query SEQR. That is, re-ordering hits based on the statistics of the hits themselves improves our ability to distinguish true hits from false hits. The implication is that if an index term (or a close neighbor) occurs in many hits, it is a better indicator of a true-hit than an index term that is not well-represented in the hits. We have referred to SEQR run in this way as 1.5-query SEQR.

This method has the advantage that it is almost as fast as 1-query SEQR, and performs nearly as well as 2-query SEQR in the high-threshold region of the ROC curves. (See the results section). However, additional computation is needed to run this method over 1-query SEQR, due to the need to recalculate the Jaccard index for the each hit, rather than directly using the scores returned from Solr.

## Creating a sequence index

Preparation of the sequence database proceeds straightforwardly using the methods described above. Each sequence is examined in turn and an indexing term is generated for each position in the sequence by converting each n-mer to a cluster index. These indexing terms are fed into a standard search engine which holds the indexing terms as inverted indices. The inverted index maps the indexing terms back to the sequences where they are found.

To query the sequence database, a query sequence is also transformed into a set of index terms in the same manner as the database sequences and the index terms are used to search the database. Each index term in the database points to a set of protein sequences.  The number of matching terms between a query and a database sequence is the number of times the index terms from a query sequence retrieve a particular database sequence from the inverted indices.  The score used to order the results can be selected from a variety of standard candidates, including the Jaccard index.

## ROC analysis

We used ROC analysis to decide on the optimal algorithm and best parameters. ROC curves plot sensitivity vs specificity at different confidence levels by plotting the number of true-positives vs the number of false positives at each confidence level.  We made ROC curves for different sets of SEQR parameters and algorithm configurations.

The database we used to index and search is Swiss-Prot (UniProt) with 460,111 records.

We selected 300 sequences at random from Swiss-Prot to serve as our test set.  Of these, we over-represented human sequences by selecting 100 human sequences at random, and added 200 non-human and non-virus sequences.  Human sequences were over represented as they are common search queries.

To establish a standard of truth, we searched Swiss-Prot with these same sequences using DELTA-BLAST (ref) using an e-value of 0.01, and allowing for up to 10,000 hits per query.  We set these search parameters permissively so that when we compare SEQR hits to our true-positives, we avoid missing borderline cases.

To compare ROC curves, the area under the curve (AUC) for the first 200,000 false positives was calculated.

## Scoring hits

We use a Jaccard index to judge the goodness of query-to-hit matches:

$$J(A, B) = \frac{|A \cap B|}{|A \cup B|} = \frac{|A \cap B|}{|A| + |B| - |A \cap B|}$$

The intersection of A and B is the number of index terms a query and hit have in common.  The union of A and B is the number of index terms in the query, plus the number of index terms in the hit, less the number of index terms the 2 have in common. If an index term from the query is found multiple times in a database sequence, the match is only counted one time in the intersection. This avoids over-weighting repeats. The range of the Jaccard index is from 0 to 1, with 1 being an exact match. The Jaccard index is also referred to as a Tanimoto score.

For our work with 2-query SEQR, the definition of the Jaccard index is slightly modified because we allow cluster-to-neighbor matches. If an index term from a hit matches either an index term in the 2nd query, or one of its cluster neighbors that was used in the 2nd query, then this counts as a matching term. Once a position in the query is counted as a matching position, it is not counted again, as this would over-weight repeats in the hit.

Also, for 2-query SEQR, the number of index terms in the hit is the same. But the number of index terms in the 2nd query is the number of positions used in the 2nd query. So if an index term and one or more of its cluster neighbors are used in the 2nd query, this counts as 1 position in the query when calculating the Jaccard index.

## Software Used

The indexing software was written in C++. The search engine used was Apache Solr version 4.10 (Apache). Solr, written in Java, was modified to return the Jaccard index, also known as the Tanimoto similarity, as the score for a hit.

# Results

Using the methods outlined above, we measured the effect of various algorithm configurations and parameters on SEQR search results in comparison to DELTA-BLAST.

## Average PSSM vs seed PSSM

In the SEQR algorithm, sequences are indexed using clusters of n-mers, where each cluster consists of an averaged PSSM as well as a seed PSSM. We have found that sensitivity and specificity are improved when we compare n-mers to the average PSSM of each cluster, rather than the seed PSSM of each cluster, when making the clusters. This means that allowing the centroids of the PSSMs of the clusters to drift during the clustering procedure results in improved performance. For these tests, the threshold was varied to determine the best threshold for each case. We also compared the algorithm to an exact-match algorithm where each cluster consisted of one unique n-mer. Both clustering methods had significantly better results over the full range of the ROC curve than the exact-match algorithm (figure 2).

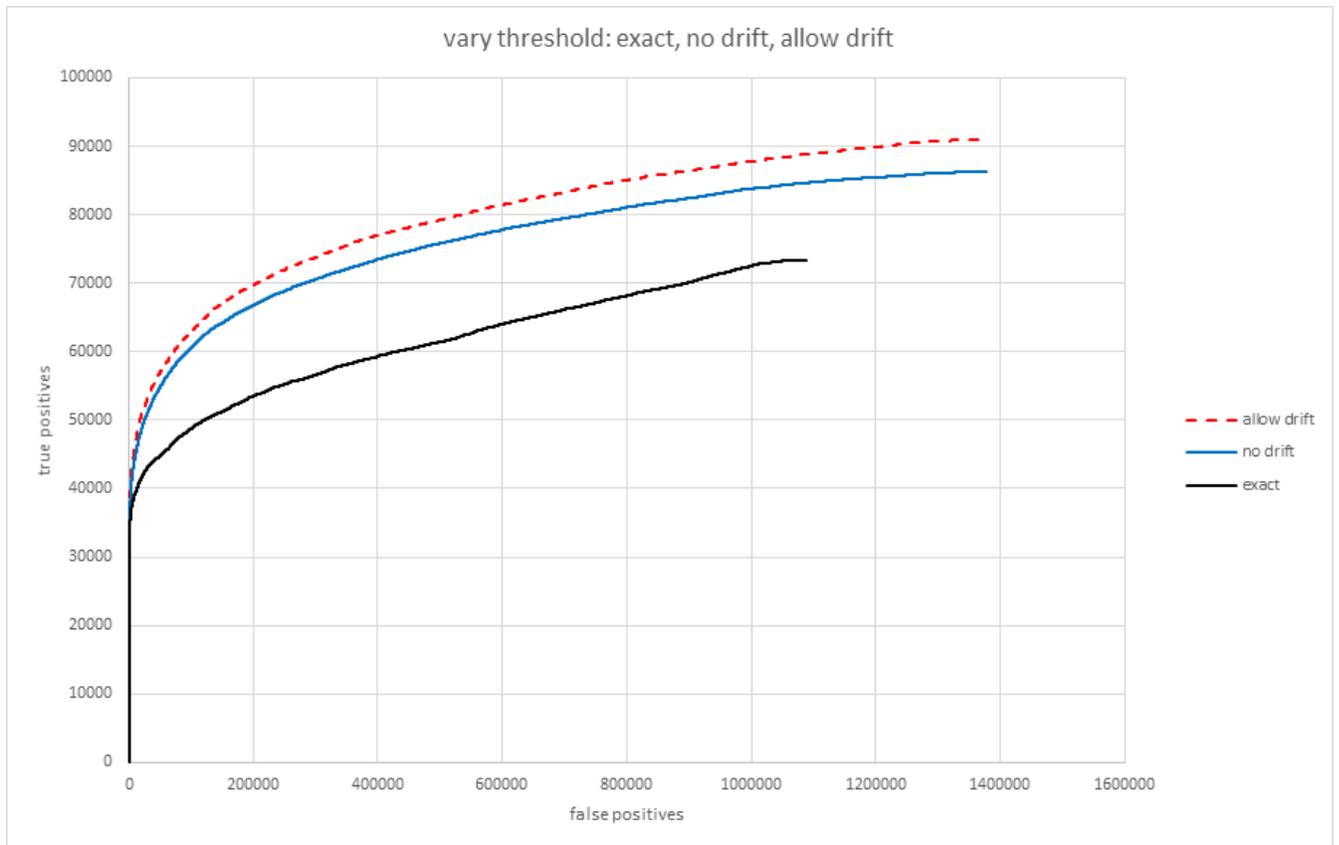

**Figure 2.** These ROC curves show that allowing the centroids of the PSSMs of the clusters to drift during the clustering procedure (red dashed) is better than using fixed PSSMs (blue solid). For reference, the black solid line is when no clustering is done and each n-mer is assigned to a unique index. These curves were made for 5-mers using the BLOSUM62 matrix. Thresholds were varied to maximize the area under the curves. The curves with the highest area-under-curve at a limit of 200k false hits are shown. [See fig. 2s in the supplemental materials for the area under each ROC curve we tested].

## Best hit vs. first-hit vs. fast-cluster

We experimented with adding n-mers to the first similar cluster in a set of clusters, adding n-mers to the most similar cluster in the set, and a fast-clustering method which adds n-mers to the most similar cluster in the list but saves time by avoiding most of the query to cluster comparisons. We found that it is advantageous to compare an n-mer to all clusters in the list before assigning it to its best-matching cluster, and that our fast-clustering method is comparable to adding n-mers to their best hits. See the ROC curve in figure 3 for a comparison of the 3 methods.

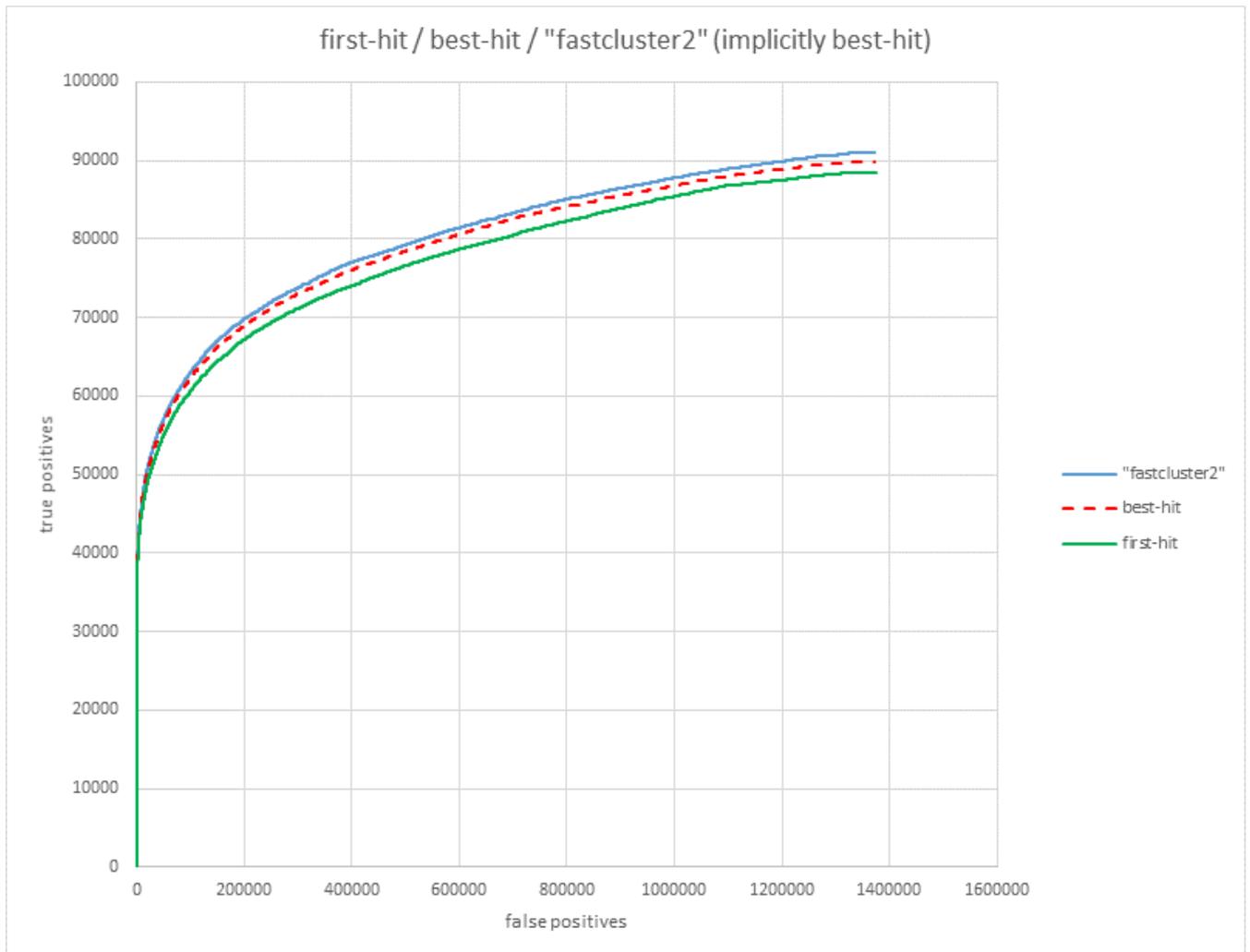

**Figure 3.** These ROC curves compare the best-hit method (red-dashed), first-hit method (green solid), and the fast-clustering method (blue solid). The best-hit method and the fast-clustering method (which also implicitly uses best-hit comparisons) performed better than the first-hit method. [See figs. 3s1 and 3s2 in the supplemental materials for more ROC curves, and the area under each ROC curve we tested].

## Reassigning n-mers

We experimented with reassigning n-mers once the full set of clusters was determined. In this way, each n-mer is tested against all clusters, and not just those that are present when the n-mer is first added to the growing list of clusters. We found that up to 8 or so rounds of reassigning n-mers to their best-matching clusters gives modest performance improvement (figure 4).

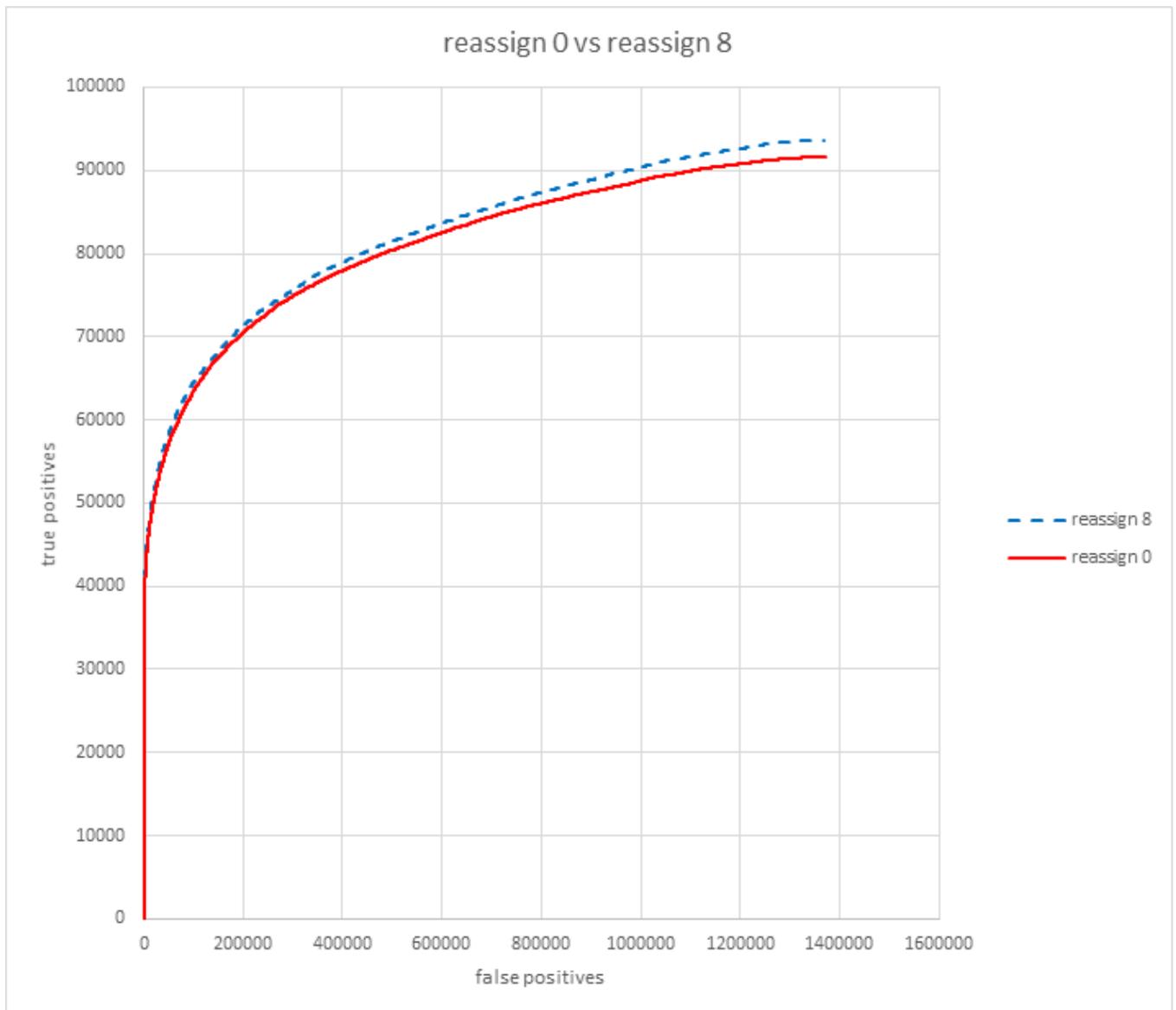

**Figure 4.** These 2 ROC curves show that there is modest improvement of SEQR when n-mers are reassigned to their best-matching clusters following the initial clustering. These clusters were made with the fast clustering method, 5-mers, using the BLOSUM62 substitution matrix, and allowing the PSSM to drift during clustering. [See figs. 4s1 and 4s2 in the supplemental materials for the area under each ROC curve, and the number of n-mers that were reassigned during each round of reassignment].

## Testing Substitution Matrices

We tested clustering n-mers using 8 different standard substitution matrices (BLOSUM[45|50|62|80|90], and PAM[30|70|250]).  The set of clusters we made using BLOSUM62 gave us the best performance at accurately retrieving hits from Swiss-Prot, with BLOSUM80 and BLOSUM90 not far behind (figure 5).  For these cluster-comparison tests, we varied the threshold when making the clusters such that there were roughly 200,000 clusters in each set (+/- 5%) to mitigate issues with different thresholds for each substitution matrix.

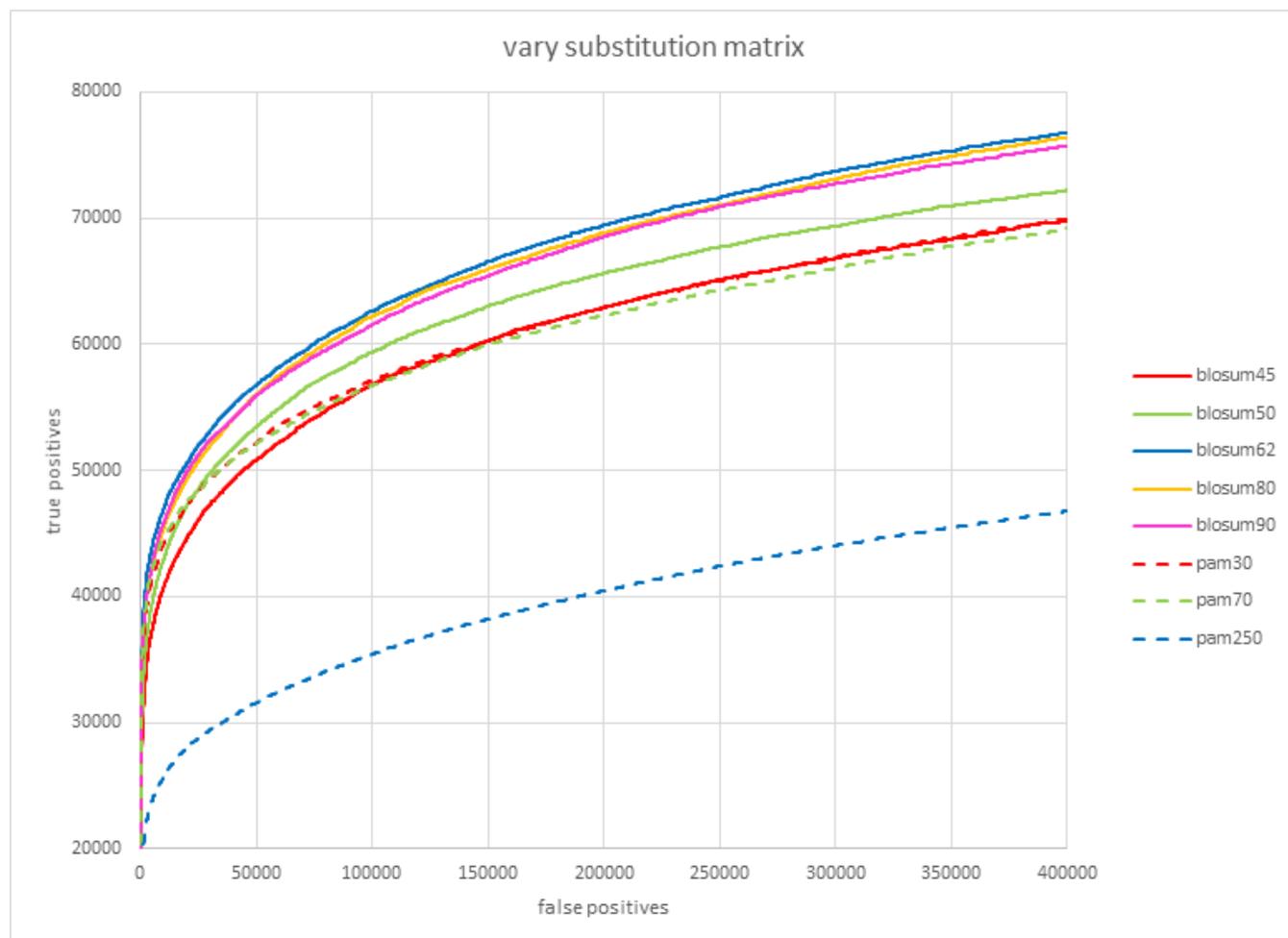

**Figure 5.** These ROC curves were made for 8 sets of clusters, each constructed using a different standard substitution matrix. The threshold for each set was chosen so that there were about 200,000 clusters in each set. Fast clustering was used to make the clusters, the PSSMs were allowed to drift, and 5-mers were used in each case. The best curve in terms of sensitivity and specificity is for BLOSUM62 (blue-solid), as judged by the area under curve to 200k false positives. Note that the origin for true-hits does not start at 0. [See fig. 5s in the supplemental materials for the area under each ROC curve we tested].

## Testing n-mer size

We tested using 3-mers, 4-mers, 5-mers, and 6-mers as the basis for our index searches. We found that 5-mers and 6-mers perform nearly equally, with the slight edge to 5-mers in the early part of the ROC curves (see figure 6).  For each n-mer size, the cluster threshold was varied, and the optimal threshold was chosen as judged by the area under the corresponding ROC curves.

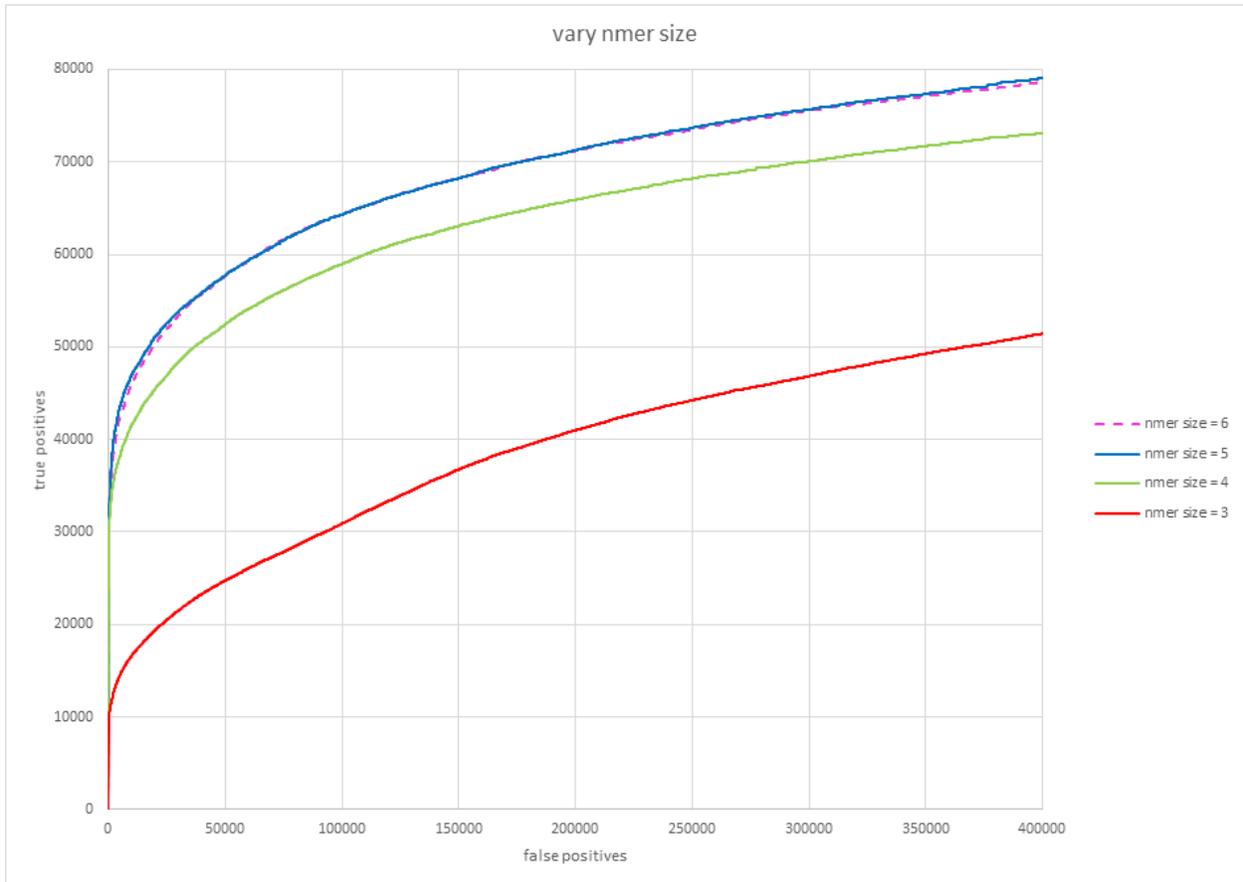

**Figure 6.** These 4 ROC curves are for n-mers clusters made from 3-mers, 4-mers, 5-mers, and 6-mers. The sensitivity and specificity of the curves improved as the n-mer size was increased from length 3 to 5.  6-mers performed slightly worse than the 5-mers, most noticeably in the early part of the curves.  For each n-mer size, the threshold was varied to optimize the area-under-curve, while other input parameters were held constant.  In each case, BLOSUM62 was used, the n-mers were compared to the average PSSM of the clusters, fast clustering was used, and 10 iterations of reassigning n-mers to clusters were performed. [See fig. 6s in the supplemental materials for the area under each ROC curve we tested].

## Varying the threshold

A key parameter of the algorithm is the threshold for inclusion of an n-mer into the cluster. To set this parameter, we used optimal conditions for making clusters as determined in the previous sections. Namely, for this test, we compared each n-mer to the average PSSM of the cluster, used the fast-clustering method which compares n-mers to all potential PSSMs and adds them to their best-matching cluster, used the BLOSUM62 matrix, used n-mers of size 5, and reassigned n-mers 10 times following clustering. This test indicated that threshold = 3.6 is near optimal (fig. 6).

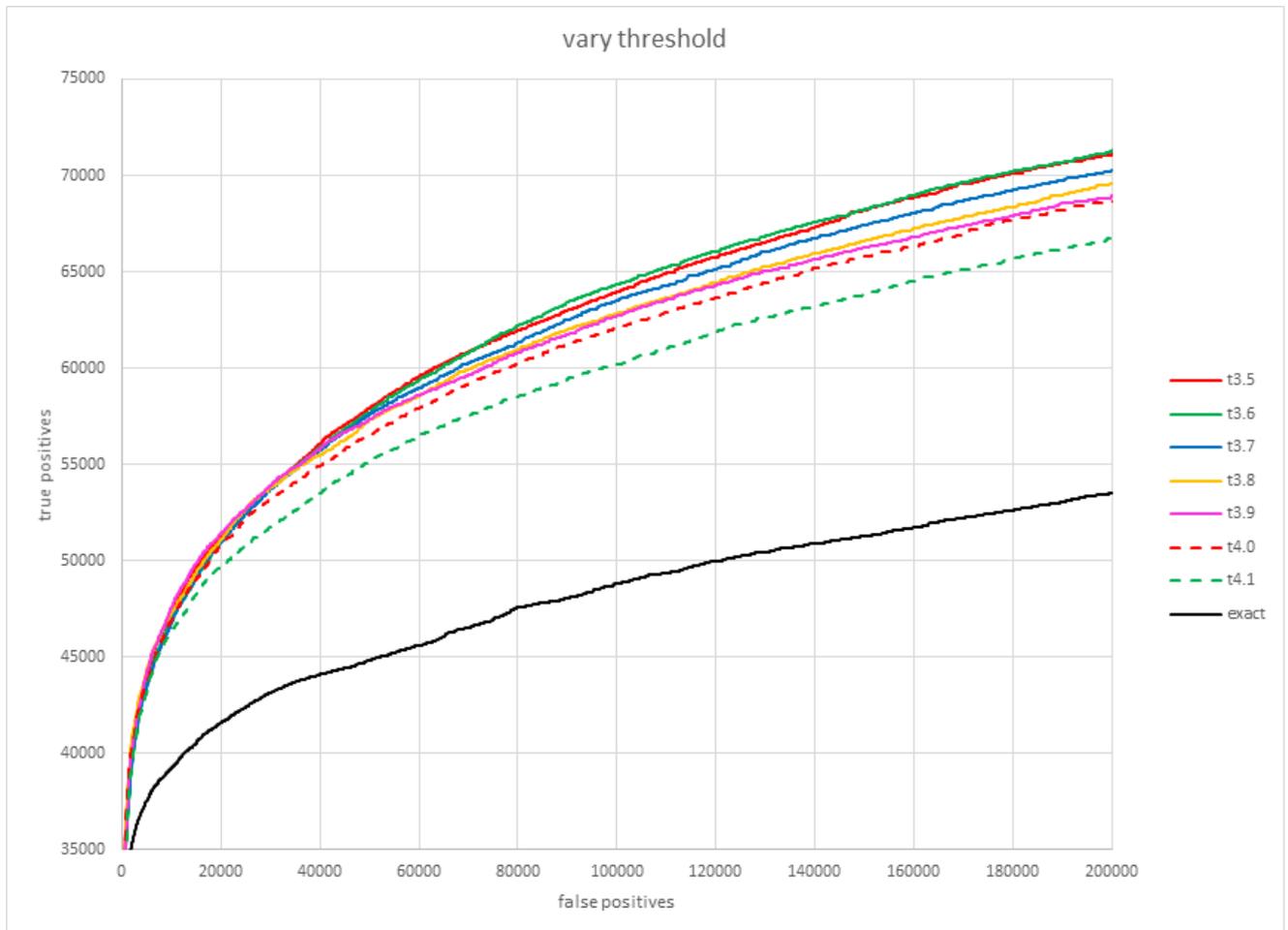

**Figure 7.** All ROC curves in this set used the same clustering conditions (n-mer=5, allow-drift, best-hit, reassign=10), except the threshold was varied between 3.5 and 4.1 in steps of 0.1. The exact-match results are also shown in black. The area-under-curve for threshold = 3.6 (green) is the best for the portion of the ROC curves below 200k false positives. Note that to show better separation of the curves, the origin of the y-axis (true positives) is not shown, and the x-axis (false positives) is zoomed in, compared to the other figures. [See fig. 7s in the supplemental materials for the area under each ROC curve we tested].

## Conversion of Jaccard index to percent identity

We used the Jaccard index in this work to rank hits but this measure has limited intuitive value to users. On the other hand, percent identity is commonly used and intuitive when comparing 2 sequences. Therefore, we apply a Jaccard index → %id transformation in order to display %id with each hit. To make a function that transforms a Jaccard index to %id, we made a plot of %id vs Jaccard index for all SEQR hits of 294 query sequences and fit a log function to the data (figure 8). The points are plotted for the SEQR hits that are also DELTA-BLAST true positives using %id retrieved from DELTA-BLAST and the Jaccard index from the SEQR search. The function is constrained so that the estimated %id for a Jaccard index of 1 is 100, and the estimated %id for a Jaccard index of 0 = 0.

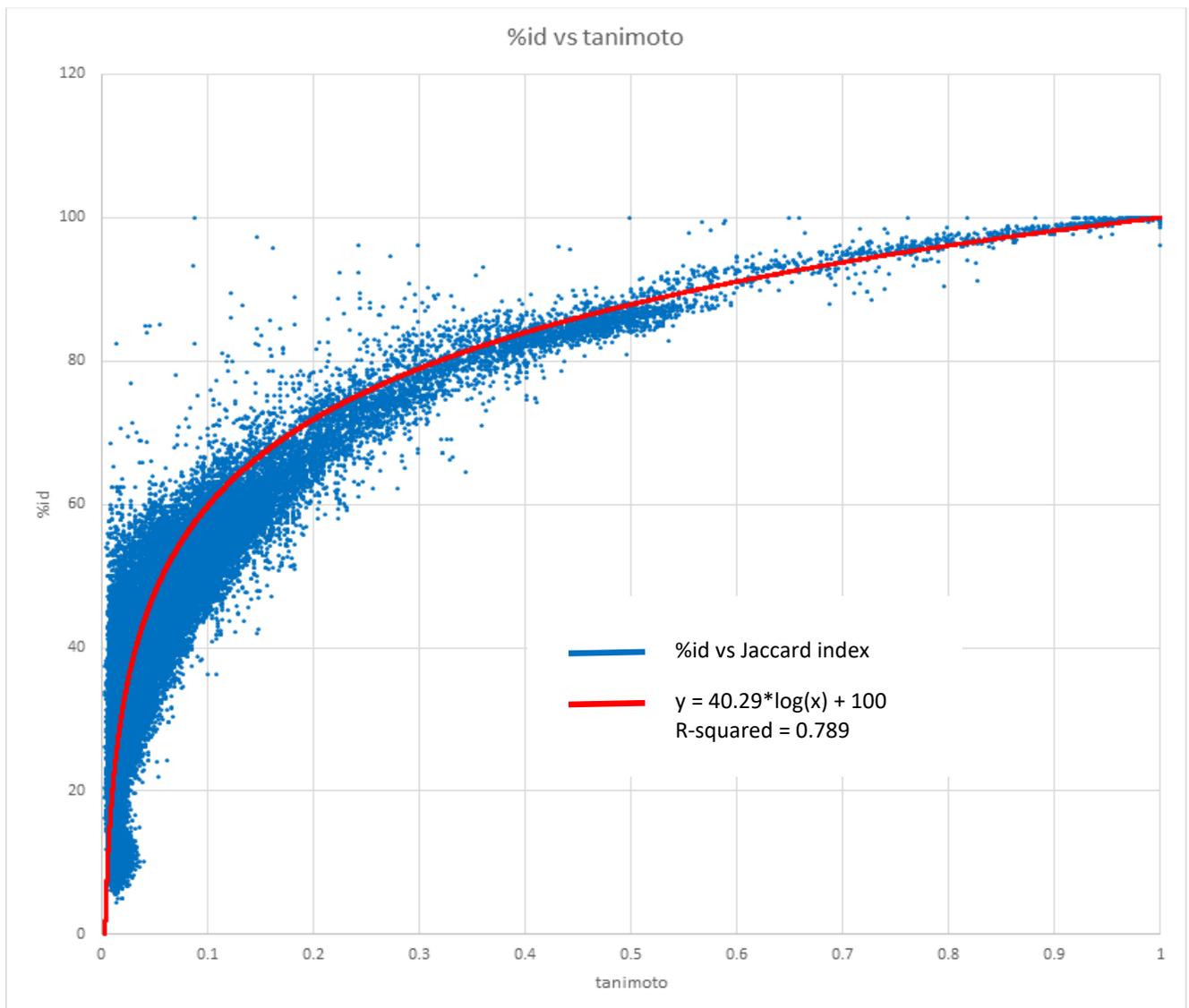

**Figure 8.** Here is the plot of %id vs Jaccard index for all SEQR hits of 294 query sequences where a DELTA-BLAST true-positive result was also found. The red line is for y = 40.29 * log(x) + 100 which maximizes R-squared given the constraint that the %id of the endpoints are fixed at 100 and 0 when the corresponding Jaccard indices are 1 and 0. R-squared = 0.789.

## 2-query SEQR

We performed a series of ROC tests to determine the best conditions for running 2-query SEQR. In particular, we varied the threshold for clustering n-mers, the cluster-neighbor threshold for deciding which cluster-neighbors to use in the $2^{nd}$ query, and the number of query terms to use in the $2^{nd}$ query as a fraction of the number of query terms used in the $1^{st}$ query. The cluster-neighbor threshold was varied relative to the cluster threshold so we refer to this as a delta-threshold.

In the first set of tests, we considered 5 clustering thresholds (3.3, 3.5, 3.7, 3.9, 4.1), 4 cluster-neighbor deltas (-0.6, -0.4, -0.2, -0.0), and 8 $2^{nd}$-query fractions (0.1, 0.2, 0.4, 0.7, 1.0, 1.5, 2.0, 3.0) for a total of 5 x 4 x 8 = 160 ROC curves.  In this case, the best conditions were for a threshold = 4.1, delta-threshold = -0.6, and the $2^{nd}$-query fraction = 2.0 or 3.0.  In the $2^{nd}$ set, we focused on this better performing zone and considered 5 thresholds (3.7, 3.9, 4.1, 4.3, 4.5), 6 cluster-neighbor deltas (-1.4, -1.2, -1.0, -0.8, -0.6), and 5 $2^{nd}$-query fractions (0.5, 1.0, 1.5, 2.0, 3.0) for an additional 5 x 6 x 5 = 150 ROC curves.  The best parameters are for threshold = 4.3, neighbor-threshold = 3.3 (delta-threshold =-1.0), and a $2^{nd}$-query fraction = 3.0.  The best 2-query ROC curve is shown below, along with the best 1-query, 1.5-query, and exact-match ROC (see figure 9).

## 1.5-query SEQR

1.5-query SEQR attempts to use the techniques of 2-query SEQR but without requiring the second search. This is done by running 1-query SEQR as usual, creating a $2^{nd}$ query as we would in 2-query SEQR (but not querying Solr with it), then rescoring the hits from the $1^{st}$ query against the new query. This algorithm gives better performance than 1-query SEQR in terms of search specificity and selectivity, while avoiding the speed penalty of running the search a $2^{nd}$ time. However, there is a small cost to pay for the improved performance: the Jaccard index of each hit needs to be recalculated since each index term in the hit can now match either an index term of the query, or one of its cluster neighbors.

To optimize parameters for running 1.5-query SEQR, we used 2-query SEQR results as a guide. Therefore, we focused on 4 cluster-thresholds (3.9, 4.1, 4.3, 4.5), 4 cluster-neighbor deltas (-1.4, -1.2, -1.0, -0.8), and 4 $2^{nd}$ query fractions (1.0, 1.5, 2.0, 3.0) to create a total of 64 ROC curves.  In contrast, we made a total of 310 ROC curves when determining optimal parameters for 2-query SEQR. Our best results were for a threshold of 4.1, a cluster-neighbor threshold of 3.1, and a $2^{nd}$ query fraction of 3.0 times as many hits in the $2^{nd}$ query as the original query.

These results show that by using statistics from the hits of a query, we can substantially improve our ability to distinguish true-positives from false positives, as evidenced by the ROC curves in figure 9. At the beginning of the ROC curve, which corresponds to the start of the hit list for each algorithm, 1.5-query SEQR outperforms 1-query SEQR. It does not outperform 2-query SEQR.

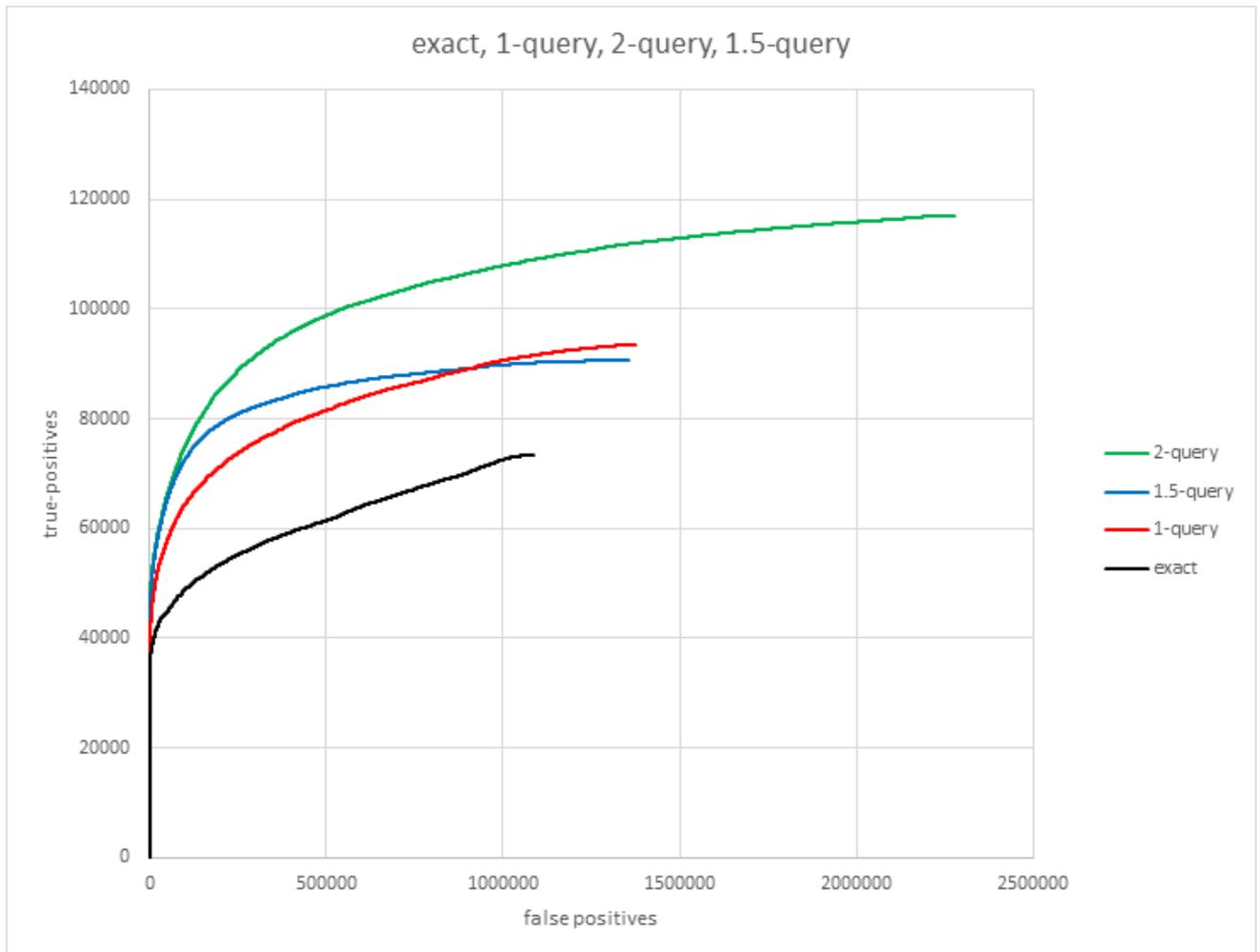

**Figure 9.** These ROC curves show the improvement of optimized 2-query SEQR (green) and 1.5-query SEQR (blue) over our best 1-query SEQR (red) and exact-match indices (black) at retrieving hits for 294 random queries and distinguishing true-positives from false-positives. The 2-query SEQR curve is for clusters that were made with a threshold of 4.3, and using neighbor clusters down to a threshold of 3.3 when making the 2nd query. The 1.5-query SEQR curve was made with a threshold of 4.1, and a cluster-neighbor threshold of 3.1. Both 2-query and 1.5-query SEQR used 3.0 times as many index terms in the 2nd query as were used in the 1st query. The red curve is for our best 1-query SEQR, where clusters were made with a threshold of 3.6, using 5-mers, allowing the PSSM of the clusters to drift, using the fast-clustering method, and reassigning n-mers to clusters 10 times. These same conditions (aside from the threshold) are also used when making the 1st query of the 1.5-query and 2-query methods. [See figure 9s1 and 9s2 in the supplemental materials for the area under each ROC curve we tested].

## Timing Tests

The principle benefit of searching with Solr over BLAST is that searches can run much faster since they use indices on the sequence database.

We observe that querying a Solr database that has 52.5M sequence records with our test set of 294 query sequences takes 2748 seconds when using 9 shards of Solr on an Intel Xeon machine running at 2.50 GHz. (This corresponds to using 9 cores of the machine). A comparable BLASTP search with 30 query sequences of the same size database running on 9 threads on the same machine takes about 2485 seconds.  SEQR, therefore, processes requests at a rate of 2748 sec / 294 queries = 9.35 secs/query, while BLASTP processes requests at a rate of 2485 sec / 30 queries = 82.84 secs/query, giving SEQR about a 9-fold speed improvement over BLASTP.

## Comparison to BLAST

One point which may be lost in the discussions above is that SEQR is less sensitive than DELTA-BLAST, particularly when the percent identity between query and hit is less than about 25%.  In some sense, all the discussion about improving the sensitivity of SEQR is taking place around the edges.  Above 50% identity, all the SEQR methods (exact-match, 1-query, 2-query, 1.5 query) perform well.  Below 25% identity, none of the SEQR methods do very well.  But in the range of 25% - 50% sequence identity, the SEQR optimizations discussed in this paper make a difference.  Our goal is to continue increasing the sensitivity of SEQR.  Looking at the figures below, we see, for example, that 2-query SEQR finds about twice as many true-hits as exact-match SEQR in the range of 25% - 30% sequence identity.

When running our ROC analysis on 294 random query sequences, we use the SEQR hits to annotate our DELTA-BLAST hits, noting for each BLAST hit whether or not a SEQR hit was found.  In the figure below, we simply counted up the number of BLAST hits and misses, in 20 different %-identity bins, regardless of the e-value of the BLAST hit or the Jaccard index of the SEQR hit.  This was repeated for exact-match, 1-query, and 2-query SEQR.  These bins correspond to the number of true SEQR hits in each of these bins for each of 3 different methods, as well as the total of the hits and misses.  Note that this plot does not say anything about the quality score of the hits.

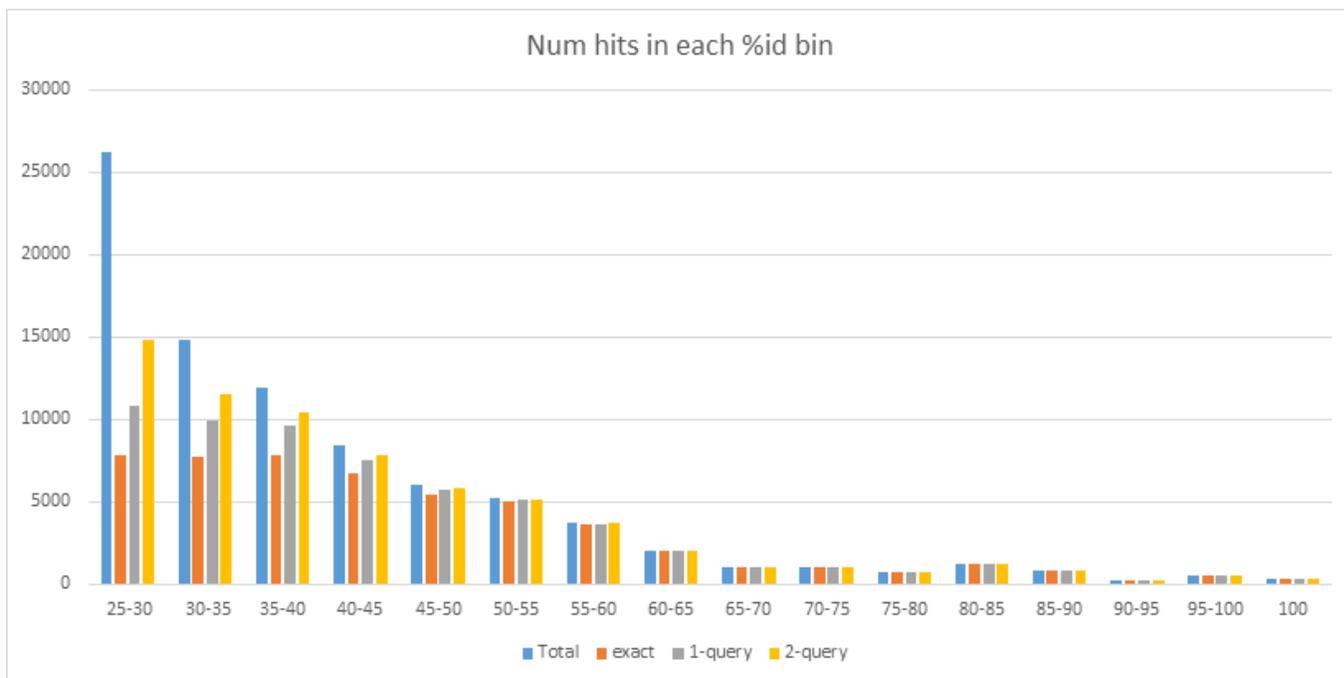

**Figure 10.** These plots show the total number of DELTA-BLAST true-positives and the number of SEQR hits for 3 different SEQR methods. The true-positives and SEQR hits are divided into 20 different %id bins as reported by DELTA-BLAST (16 bins are shown). All 3 SEQR methods do well at finding BLAST hits above 50% sequence identity, and poorly at finding BLAST hits below 25% identity. 2-query SEQR has better sensitivity than 1-query SEQR which has better sensitivity than exact-match SEQR in all bins. These results are for our best-case, 1-query, and 2-query methods.

## Conclusions

SEQR can search sequence databases significantly faster than methods such as BLAST. However, as implemented, it has lower sensitivity. We believe the low sensitivity of SEQR can be improved with better indexing terms, and that low sensitivity is not intrinsic to the algorithm. Already, SEQR performs with more sensitivity than identical k-mer searching.

When generating the clusters of n-mers that form the SEQR indexing PSSMs, we optimized several parameters. We found that it is best to compare n-mers to the average PSSM of a cluster, rather than the seed of the cluster, thereby allowing the PSSM of the cluster to drift from the seed PSSM during n-mer clustering. Additionally, performance is improved if n-mers are assigned to their best-matching cluster, rather than their first-matching cluster. Similarly, it is best to reassign n-mers to their best-matching clusters several times, once the full set of clusters is determined. Finally, the standard substitution matrix BLOSUM62 and n-mers of size 5 gave us the best performance distinguishing hits and misses.

We found the Jaccard index was useful for ranking SEQR hits and was straightforward to convert into an estimated percent identity.

Two additional variants of SEQR were developed to further improve sensitivity. One uses a $2^{nd}$ query to pull in additional hits. The $2^{nd}$ query is derived from the index terms of an initial query, plus their closely-related neighbors, that are frequently observed in the hits of the initial query. Combining the hits of these 2 queries gives improved performance over using the hits from the $1^{st}$ query alone.

While developing the 2-query method, we noted that simply constructing the $2^{nd}$ query, and scoring the hits from the $1^{st}$ query against the $2^{nd}$ query, led to improved performance over the 1-query method. This method gives almost the same sensitivity of the 2-query method without the expense of running the $2^{nd}$ query. In effect, this method reorders the hits of the $1^{st}$ query using statistics from the query.

We have developed a public service using SEQR. It is at https://www.ncbi.nlm.nih.gov/Structure/seqr. This service demonstrates the incorporation of SEQR into a general purpose search engine, allowing users to subset by various types of indices, such as taxonomy, and to do free text searching. These additional tools allow users to find sequences of interest by using combinations of queries for similar sequences and on other indices.

# Future work

## Sensitivity improvements

The ROC analysis shows that the sensitivity of the algorithm is the same as DELTA-BLAST down to 40% sequence identity, a level which is useful for many real world applications that do not require traversing large evolutionary distances. The lack of sensitivity in the twilight zone of sequence similarity is not an intrinsic feature of the algorithm, however, but of the indexing PSSMs. In general, PSSMs are a powerful method to deduce distant similarities as demonstrated by algorithms like PSI-BLAST and RPS-BLAST [cite]. If the current indexing PSSMs can be replaced with more sensitive PSSMs, such as those found in Pfam [cite] or the Conserved Domain Database [cite], the sensitivity of the algorithm may approach that of DELTA-BLAST.

## Extension to searching DNA

It is straightforward to convert the current algorithm into a version that takes a DNA sequence and translates all six frames into protein sequences, which are then used to search a protein database. Searching a DNA sequence database by DNA directly requires incorporating locality when the database contains large DNA sequences, such as chromosomes.

# Acknowledgments

We are indebted to David J. Lipman, Stephen H. Bryant, Stephen Altschul, and Alejandro Schaeffer for useful discussions. This research was supported by the Intramural Research Program of the NIH, National Library of Medicine.

UniProt within days down to 20%-30% maximum pairwise sequence identity. kClust owes its speed and sensitivity to an alignment-free prefilter that calculates the cumulative score of all similar 6-mers between pairs of sequences, and to a dynamic programming algorithm that operates on pairs of similar 4-mers. To increase sensitivity further, kClust can run in profile-sequence comparison mode, with profiles computed from the clusters of a previous kClust iteration. kClust is two to three orders of magnitude faster than clustering based on NCBI BLAST, and on multidomain sequences of 20%-30% maximum pairwise sequence identity it achieves comparable sensitivity and a lower false discovery rate. It also compares favorably to CD-HIT and UCLUST in terms of false discovery rate, sensitivity, and speed. CONCLUSIONS: kClust fills the need for a fast, sensitive, and accurate tool to cluster large protein sequence databases to below 30% sequence identity. kClust is freely available under GPL at [http://toolkit.lmb.uni-muenchen.de/pub/kClust/](http://toolkit.lmb.uni-muenchen.de/pub/kClust/).

Henikoff, S. and J. G. Henikoff (1992). "Amino acid substitution matrices from protein blocks." <u>Proc Natl Acad Sci U S A</u> **89**(22): 10915-10919.

Methods for alignment of protein sequences typically measure similarity by using a substitution matrix with scores for all possible exchanges of one amino acid with another. The most widely used matrices are based on the Dayhoff model of evolutionary rates. Using a different approach, we have derived substitution matrices from about 2000 blocks of aligned sequence segments characterizing more than 500 groups of related proteins. This led to marked improvements in alignments and in searches using queries from each of the groups.

Kato, T. (1992). <u>Database architecture for content-based image retrieval</u>.

This paper describes visual interaction mechanisms for image database systems. The typical mechanisms for visual interactions are query by visual example and query by subjective descriptions. The former includes a sketch retrieval function and a similarity retrieval function, while the latter includes a sense retrieval function. We adopt both an image model and a user model to interpret and operate the contents of image data from the user''s viewpoint. The image model describes the graphical features of image data, while the user model reflects the visual perception processes of the user. These models, automatically created by image analysis and statistical learning, are referred to as abstract indexes stored in relational tables. These algorithms are developed on our experimental database system, the TRADEMARK and the ART MUSEUM.

Kent, W. J. (2002). "BLAT--the BLAST-like alignment tool." <u>Genome Res</u> **12**(4): 656-664.

Analyzing vertebrate genomes requires rapid mRNA/DNA and cross-species protein alignments. A new tool, BLAT, is more accurate and 500 times faster than popular existing tools for mRNA/DNA alignments and 50 times faster for protein alignments at sensitivity settings typically used when comparing vertebrate sequences. BLAT's speed stems from an index of all nonoverlapping K-mers in the genome. This index fits inside the RAM of inexpensive computers, and need only be computed once for each genome assembly. BLAT has several major stages. It uses the index to find regions in the genome likely to be homologous to the query sequence. It performs an alignment between homologous regions. It stitches together these aligned regions (often exons) into larger alignments (typically genes). Finally, BLAT revisits small internal exons previous paper). We note that the weighting function can find translational initiation sites within sequences that were not included in the training set.